\documentclass[aps,twocolumn,prd,nofootinbib,superscriptaddress,preprintnumbers,10pt]{revtex4-2}

\usepackage[utf8]{inputenc}

\usepackage{mathtools}
\usepackage{amsfonts}
\usepackage{amssymb}
\usepackage{mathrsfs}
\usepackage{bbm}
\usepackage{braket}
\usepackage{slashed}
\usepackage{tensor}

\usepackage{graphicx}
\usepackage{xcolor}
\usepackage[normalem]{ulem}
\usepackage{array}
\usepackage{comment}

\usepackage{placeins}
\usepackage{makecell}
\usepackage[caption=false]{subfig}
\usepackage{float}

\usepackage{xspace}
\usepackage{xfrac}
\usepackage{hyperref}
\usepackage[nameinlink]{cleveref}
\usepackage{appendix}
\usepackage{units}

\usepackage{xifthen}
\usepackage{booktabs}
\usepackage{multirow}
\usepackage{courier}
\usepackage[dvipsnames]{xcolor}
\hypersetup{
	colorlinks,
	linkcolor={blue!75!black},
	citecolor={blue!75!black},
	urlcolor={blue!75!black}
}

\newcommand{\ssection}[1]{\emph{#1.---}}

\Crefname{equation}{Eq.\!}{Eqs.\!}

\DeclareMathOperator{\Tr}{Tr}
\DeclareMathOperator{\tr}{tr}

\newcommand{\bs}[1]{{\boldsymbol{#1}}}

\graphicspath{{./figs/}}

\newcommand{\gettitle}{Learning the generating functional for variance reduction in lattice QCD}

\newcommand{\getMITAffiliation}{\affiliation{Center for Theoretical Physics, Massachusetts Institute of Technology, Cambridge, MA 02139, USA}}
\newcommand{\getIAIFIAffiliation}{\affiliation{The NSF AI Institute for Artificial Intelligence and Fundamental Interactions}}
\newcommand{\getFNALAffiliation}{\affiliation{Fermi National Accelerator Laboratory, Batavia, IL 60510, U.S.A.}}
\newcommand{\getEdinburgAffiliation}{\affiliation{Higgs Centre for Theoretical Physics, School of Physics and Astronomy,\\
University of Edinburgh, EH9 3FD Edinburgh, United Kingdom}}
\newcommand{\getBernAffiliation}{\affiliation{Albert Einstein Center, Institute for Theoretical Physics, University of Bern, 3012 Bern, Switzerland}}

\hypersetup{
    pdftitle={\gettitle},
    pdfauthor={Abbott,Fu,Hackett,Kanwar,Romero-López,Shanahan},
    pdfkeywords={lattice field theory}{lattice gauge theory}{normalizing flows}{machine learning},
    bookmarksopen=true,
    bookmarksopenlevel=2,
    bookmarksnumbered=true
}

\begin{document}

\title{\gettitle}

\author{Ryan~Abbott}
\affiliation{Physics Department, Columbia University, New York, NY 10027, USA}
\author{Yang Fu}
\getMITAffiliation
\getIAIFIAffiliation
\author{Daniel~C.~Hackett}
\getFNALAffiliation
\author{Gurtej~Kanwar}
\getEdinburgAffiliation
\author{Fernando~Romero-L\'opez}
\getBernAffiliation
\author{Phiala~E.~Shanahan}
\getMITAffiliation
\getIAIFIAffiliation

\preprint{FERMILAB-PUB-26-0404-T, MIT-CTP/6051}

\begin{abstract}
The generating functional in quantum field theory provides the natural framework for constructing correlation functions as derivatives with respect to source operators. We present a methodology that leverages machine-learned normalizing flows to reduce the variance of arbitrary $N$-point correlation functions of bosonic operators in lattice gauge field theory calculations by encoding a representation of the generating functional. We show that it is possible to systematically approach noiseless estimators of correlation functions in this framework.
We demonstrate this methodology with applications to calculations of glueball correlation functions and Wilson loops in Quantum Chromodynamics and Yang-Mills theory. The results show up to three orders of magnitude variance reduction.
\end{abstract}

\maketitle

\ssection{Introduction}
Lattice gauge theories provide a computational framework for systematically controllable, non-perturbative calculations in 
quantum chromodynamics (QCD). Lattice QCD is able to predict Standard Model observables in regimes where analytical methods are not applicable. Its success is evident: quantities such as the hadronic contribution to the muon anomalous magnetic moment~\cite{Aliberti:2025beg} and the strong coupling~\cite{DallaBrida:2026kuo} have been determined with unprecedented precision using lattice QCD. However, the next generation of statistical precision requires a breakthrough in the tools used for Monte Carlo estimates of observables. 
An emerging direction that may provide such a breakthrough is the application of normalizing flows~\cite{tabak2010,tabak2013,rezende2016variational} with machine-learned components to lattice field theories~\cite{DelDebbio:2021qwf,Foreman:2021ljl,Bacchio:2022vje,Tomiya:2022meu,Finkenrath:2022ogg,Abbott:2022zsh,Abbott:2023thq,Albandea:2023wgd,Bialas:2023fyj,Gerdes:2024rjk,Bonanno:2025pdp}. Their application as a tool to reduce variance of statistical estimators has shown particularly promising results~\cite{Abbott:2024mix,Abbott:2024kfc,Abbott:2026ylv,Hermansson-Truedsson:2026zlp}.

Here we take a further step and introduce a method to directly learn the generating functional for arbitrary operators using normalizing flows. The key observation is that normalizing flows can connect infinitesimally close distributions, giving noise-reduced, unbiased estimates of partition-function ratios. 
We demonstrate that these estimates can be driven to the zero-variance limit by improving model quality. Equivalently, flows provide
a class of exact control variates that modify the variance of observables while strictly preserving their mean, and that can be optimized to perfectly cancel observable fluctuations.
This may lead to a shift in perspective for lattice calculations: in the theoretical limit of perfect models, the flow approach entirely eliminates the signal-to-noise problem by directly encoding observables independent of any Monte Carlo sampling; in practice, one may hope to push the precision of observables well beyond the state of the art, even with finite parameterizations and training.

While a variety of other variance-reduction methods have been introduced for lattice field theory, this is---to the best of the authors' knowledge---the first instance of a method that can construct a family of improved estimators with the guarantee of convergence to a zero-variance limit. For example, other variance-reduction tools such as contour deformations~\cite{Detmold:2020ncp,Detmold:2021ulb,Lin:2023svo,Lawrence:2023sfc}, control variate methods based on Schwinger-Dyson equations~\cite{Bhattacharya:2023pxx,Lawrence:2024xsi}, related total derivatives~\cite{Bacchio:2023all}, or machine learned observables~\cite{Yoon:2018krb,Bedaque:2023ovz,Oh:2025fpq} have no guarantees on theoretically achievable variance reduction.
Closely related to this work, Refs.~\cite{Catumba:2023ulz,Catumba:2025ljd} propose to evaluate $n$-point functions by up to $n-1$ derivatives of the generating functional using an approach amounting to stochastic estimates of flows. By instead constructing deterministic flows, we eliminate the fundamental noise floor imposed by these stochastic estimates, enabling a theoretical zero-noise limit.

In this Letter, we present the learned generating-functional method
and apply it to Yang--Mills theory and QCD. We demonstrate that increasing model complexity can lead to greater variance reduction, achieving factors of up to 2000 with the models used here. We also find that the variance reduction is approximately volume-independent.

\ssection{The generating functional}
In lattice quantum field theory, using the Euclidean path integral framework, the generating function for $n$ source operators ${\boldsymbol{\mathcal O} = (\mathcal{O}_1, \dots, \mathcal{O}_n)}$ is defined as 
\begin{equation}
    Z_{\boldsymbol{\lambda}} = \int DU e^{-S(U) + \boldsymbol{\lambda}  \cdot \boldsymbol{\mathcal O}(U)},
\end{equation}
where $S(U)$ is the Euclidean action and the source coefficients are ${\boldsymbol \lambda \equiv (\lambda_1, \dots, \lambda_n)}$. 
We assume real-valued $\lambda_i$ and $\mathcal{O}_i$. For complex-valued operators, the real and imaginary components can be treated separately. 
For Grassmann-valued operators, the method can be applied with Wick-contracted correlators as source operators.

Arbitrary correlation functions of the source operators can be constructed using derivatives of $Z_{\bs \lambda}$
with respect to the source coefficients. For example, the correlation function of a single insertion of each operator is given by
\begin{equation} \label{eq:derivative-trick}
  \langle \mathcal O_1 \hdots \mathcal O_n \rangle  = \frac{\partial^n}{\partial \lambda_1 \hdots \partial \lambda_n} \frac{Z_{\boldsymbol{\lambda}}}{Z_0}
  ,
\end{equation}
where $Z_0 = Z_{\boldsymbol{\lambda}} \big|_{\boldsymbol{\lambda} = 0}$.
Derivatives $\partial/\partial \lambda_i$ and expectation values $\left< \dots \right>$ are evaluated at $\boldsymbol{\lambda}= 0$. 
The ratio of partition functions above can also be obtained as an expectation value under the $\bs{\lambda} = 0$ distribution, 
\begin{equation} \label{eq:reweighting}
    \frac{Z_{\boldsymbol{\lambda}}}{Z_0} = \langle e^{\boldsymbol{\lambda} \cdot \boldsymbol{\mathcal O}} \rangle.
\end{equation}
This expectation value can be estimated by reweighting Monte Carlo samples following the distribution ${r(U) = e^{-S(U)}/Z_0}$, and differentiating this estimator directly leads to the standard method of measuring correlation functions in lattice field theory. The well-known signal-to-noise problem~\cite{Parisi:1983ae,Lepage:1989hd} is closely tied to the structure of this reweighting estimator.

\ssection{Learning the generating functional}
We aim to reduce the variance of correlation functions by using a learned map, or flow, to connect the distributions and circumvent this reweighting problem~\cite{Abbott:2026ylv}.
A normalizing flow $f$ is a diffeomorphism mapping any input field configurations $U$ sampled from a prior distribution $r(U)$ to new samples $V \equiv f(U)$ following a different \emph{model} distribution $q(V)$. Differentiability and invertibility allow the resulting change of measure to be written explicitly as
\begin{equation}
    q(V) = r(U) |\det J_f|^{-1}, \quad J_f \equiv \partial f(U) / \partial(U).
\end{equation}

Here, we use such flows to produce systematically improvable estimators of arbitrary correlation functions of $n$ operators by directly encoding an appropriate generating functional, then taking $n$ derivatives as in \Cref{eq:derivative-trick}.
First, a flow $f_{ \boldsymbol{\lambda}}$ parameterized by the source coefficients $\boldsymbol{\lambda}$ is trained to approximately transform $r(U)$ to
\begin{equation}
    q(V) = r(U) |\det J_f|^{-1} \approx \frac{1}{Z_{\boldsymbol{\lambda}}} e^{-S(V) + \boldsymbol{\lambda} \cdot \boldsymbol{\mathcal O}(V)},
\end{equation}
with $V = f_{\boldsymbol{\lambda}}(U)$.
To evaluate the generating functional exactly,
we introduce the unnormalized reweighting factor, 
\begin{equation}
    \hat{w}_{\boldsymbol{\lambda}}(U) = e^{-S(V) + \boldsymbol{\lambda} \cdot \boldsymbol{\mathcal O}(V) + S(U)}|\det J_f|.
\end{equation}
It provides an unbiased estimator of the ratio of partition functions~\cite{Nicoli:2020evf}, and its derivative  gives the desired correlation function
\begin{equation} \label{eq:flow-corr-n}
    \frac{Z_{\boldsymbol{\lambda}}}{Z_0} = \langle \hat{w}_{\boldsymbol{\lambda}} \rangle
    \implies 
    \langle \mathcal{O}_1 \dots \mathcal{O}_n \rangle =
    \frac{\partial^n}{\partial \lambda_1 \hdots \partial \lambda_n} \left< \hat{w}_{\boldsymbol{\lambda}} \right>.
\end{equation}
In contrast, the \emph{normalized} reweighting factor has expectation value 1 and its derivative provides no information.
Both \Cref{eq:flow-corr-n,eq:reweighting} give the same ratio of partition functions and corresponding observables via the generating functional,
but their variances can differ significantly.

\ssection{Ideal flow limit}
A perfect flow is characterized by the output distribution exactly matching the target distribution. 
Importantly, the unnormalized reweighting factors in this limit provide a \emph{zero-noise} evaluation of the ratio of partition functions, i.e., on each configuration
 \begin{equation}
    \hat{w}_{\bs \lambda}(U) \xrightarrow{\substack{\mathrm{perfect} \\ \mathrm{flow}}} \frac{Z_{ \bs \lambda}}{Z_0}.
\end{equation}
In this case, \Cref{eq:flow-corr-n} evaluates the correlation function with zero variance. Such a perfect flow is guaranteed to exist under mild assumptions, such as connectedness of the space~\cite{brenier1991polar}.
This result is central to this Letter.

For imperfect flows,
the weights $\hat{w}_{\boldsymbol{\lambda}}(U)$ will always have some non-zero noise.
For a one-point function, estimated
according to \Cref{eq:flow-corr-n}, the variance can be related to the quality of the flow~\cite{Abbott:2026ylv} as
\begin{equation} \label{eq:var-ess}
    \mathrm{Var}[\partial_\lambda \hat{w}_\lambda] =\lim_{\lambda \to 0} \frac{1}{\lambda^2} \left( \frac{1}{\rm ESS} -1\right),
\end{equation}
where the effective sample size (ESS) is a standard metric for model quality in normalizing flows, defined as
${\mathrm{ESS} = \langle \hat w_\lambda \rangle^2 / \langle \hat w_\lambda^2 \rangle \in [0,1]}$. 
For one-point functions, \Cref{eq:var-ess} determines the quality of flow needed to obtain a given variance. For other cases, the improved variance is not directly predicted by the ESS, and in this work we instead directly measure the variance in each case.

We contrast this with recently explored methods to insert an operator $\mathcal{O}$ into expectation values using a flow $f_\lambda$. For a product of operators $\mathcal{A} = \mathcal{O}_1 \dots \mathcal{O}_{n-1}$
applying a derivative with respect to $\lambda$ leads to the equality
\begin{equation} \label{eq:single-der-trick}
    \langle \mathcal{A} \mathcal{O} \rangle = \langle \partial_\lambda [\mathcal{A}(V) \, \hat{w}_\lambda ] \rangle.
\end{equation}
This one-derivative trick can be recursively applied until a one-point function is reached.
While this approach has been shown to significantly reduce the variance~\cite{Catumba:2023ulz,Catumba:2025ljd,Albergo:2026fwx,Abbott:2026ylv}, the contribution from $\mathcal{A}(V)$ fluctuates even in the perfect flow limit,
meaning there is neither the guarantee of a zero-variance limit nor a definite change in the scaling of the signal-to-noise with operator separation.

\ssection{Parameterizing the flow by expansion}
Explicitly parameterizing the flow $f_{\boldsymbol{\lambda}}$ by the $n$ source coefficients may be challenging. As derived in the End Matter, a solution is to expand up to first order in all parameters,  
which is sufficient to exactly evaluate the needed first-order derivatives. This replaces the general flow $f_{\bs\lambda}$  with $2^n - 1$
``flow fields''---algebra-valued objects encoding contributions to the flow transformation at each order.
The flow fields have well-defined training objectives given in terms of the $\boldsymbol{\lambda} = 0$ distribution. Propagating derivatives in \Cref{eq:flow-corr-n} into this definition of the flow gives an explicit, zero-mean control variate with all quantities computed from samples from the $\boldsymbol{\lambda} = 0$ distribution.

A related challenge is that this method requires training flows that are specialized to a particular set of operators. For a two-point function, for example, it would require a separate flow for each operator type and source-sink separation.
We argue here that the series expansion of the flow leads to a useful approximation that avoids this issue by truncating at the linear order in the expansion, in the case of well-separated operators.
This truncation can be framed in terms of $n$ independent flows, where
the $i$th flow $f^{(i)}_{\lambda_i}$ is optimized so that
\begin{equation}
    q_i(V) \approx \frac{1}{Z_{\lambda_i}} e^{-S(V) + \lambda_i \mathcal{O}_i(V)}, \quad Z_{\lambda_i} = Z_{\bs\lambda}|_{\lambda_{j\neq i} = 0},
\end{equation}
and the combined flow is defined as
\begin{equation} \label{eq:fcomposed}
    f_{\boldsymbol{\lambda}}(U) \equiv f^{(1)}_{\lambda_1} \circ \dots \circ f^{(n)}_{\lambda_n} (U).
\end{equation}
This factorized structure allows a finite set of flows to improve a combinatorial set of correlation functions.

Time-translation invariance and other symmetries can also be utilized to relate individual flows, avoiding retraining for distinct locations on the lattice or different orientations of the operator. 
For instance, if  $f^{(i)}_{\lambda_i}$ inserts the operator $\mathcal O_i(t=0)$ and $T(t)$ is a time-translation by $t$, the translated flow $f^{(i)}_{\lambda_i}(t) =  T^{-1}(t)\circ f^{(i)}_{\lambda_i} \circ T(t)$ will insert $\mathcal O_i(t)$ at timeslice $t$.
The factorized flow approximation is particularly useful when the operators have a vacuum expectation value and the flows are imperfect, as it ensures exact cancellation of the leading $O(\lambda)$ effects, as is the case in the two-point function example below.

\ssection{Applications to Yang--Mills}
The plaquette---a $1 \times 1$ Wilson loop---is the simplest gauge-invariant operator that can be studied in the Yang--Mills theory.
Despite their simplicity, general Wilson loops are used to determine phenomenologically relevant observables, such as static quark potentials related to the string tension~\cite{Necco:2001xg} and the strong coupling~\cite{Kitazawa:2016dsl,Gockeler:2005rv}.
The plaquette can also be used as an interpolating operator in glueball spectroscopy; see below.
To illustrate the learned generating-functional methodology, we construct flows that introduce the operator
\begin{equation}
    \mathcal{O} \equiv \mathcal{G}(t) = \sum_{x} {\rm Re }  \Tr P_{12}(x, t)
\end{equation}
on individual timeslices $t$, where $P_{12}(x,t)$ indicates the plaquette in the $(1,2)$ plane with corner placed at $(x,t)$.

In the following numerical examples, we use an architecture consisting of gauge-equivariant masked residual layers~\cite{Abbott:2026ylv,Abbott:2024kfc},
such that the links are transformed as
\begin{equation}
    U'_\mu(x) = e^{\lambda F(x)} U_\mu(x),
\end{equation}
where $F(x)$ is an algebra-valued ``force'' given by a linear combination of untraced loops that begin and end at $x$. 
These models have two main hyperparameters~\cite{Abbott:2024kfc}.
The ``mod $N$'' sparsity of the variable partitioning, or ``masking pattern'', indicates what fraction of the links are transformed by a layer.
The number of gauge-equivariant convolutions, $n_{\rm pt}$, indicates how many times the frozen links are sequentially convolved together, which sets the limit on the size of loops entering the definition of $F(x)$.
Sparser masking patterns (larger $N$) and more convolutions (larger $n_{\rm pt}$) correspond to more expressive models.

\begin{figure}[b!]
    \centering
    \includegraphics[width=\linewidth]{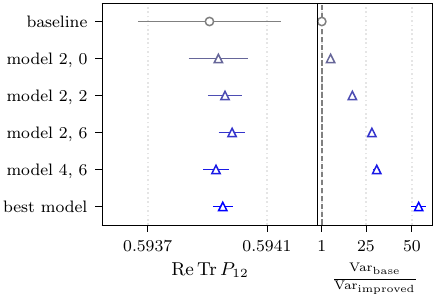}
    \caption{ Left panel: plaquette measured with the standard estimator and with different models on a pure-gauge ensemble at $\beta=6.0$ and with lattice geometry $8^3 \times 32$. Right panel: the corresponding variance reduction. Model complexity increases from top to bottom. Model labels indicate masking sparsity $N$ and number of gauge-equivariant convolutions $n_{\rm pt}$, as described in the main text.
    The ``best model'' has two stacks of layers,  ``mod 2'' and ``mod 4'', and $n_{\rm pt}=6$.
    }
    \label{fig:Wloopcompare}
\end{figure}

Using the training approach of Ref.~\cite{Abbott:2026ylv}, we train several flows with different hyperparameter choices to learn the generating functional for $\mathcal{G}(0)$, the operator at timeslice $0$.
As in Ref.~\cite{Abbott:2026ylv}, flows are trained at a small spatial volume and transferred to a larger volume. This leads to no statistically significant degradation of the model quality, while reducing training costs.

\begin{figure*}[t!]
    \centering
    \includegraphics{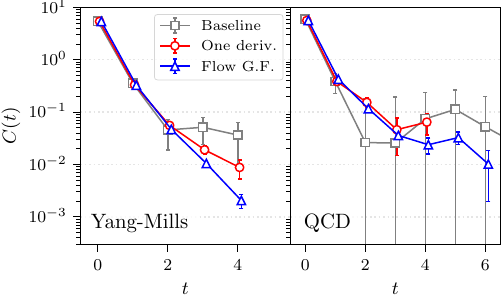}
    \hspace{0.2cm}
    \includegraphics{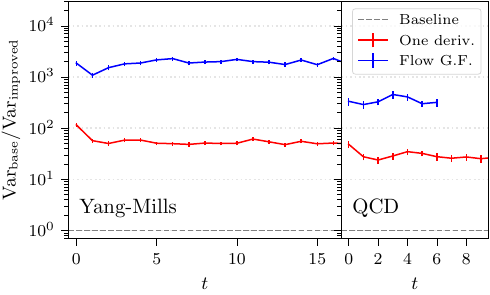}
    \caption{
    Left: Correlators $C(t)$ obtained with the standard estimator (gray squares), the flow-improved one-derivative trick~\cite{Catumba:2025ljd,Abbott:2026ylv} (red circles), and the learned generating functional method (blue triangles). All points shown are at most $2.5\sigma$ apart. Right: Achieved variance reduction, defined as the ratio between improved and baseline variance.  Panels within each figure show results for Yang-Mills at $\beta = 6.0$ and for QCD at $\beta=5.6$ and $\kappa=0.1577$, corresponding to $M_\pi \sim 600$ MeV, using ensembles with 21,000 and 1,000 configurations, respectively. In both cases, a lattice geometry of $16^3 \times 32$ was used.
    }
    \label{fig:results}
\end{figure*}

We compute the flow-improved estimate using a symmetric finite-difference approximation to the derivative in \Cref{eq:flow-corr-n} in the one-point function case,
\begin{equation}
    \langle \mathcal{O} \rangle = \frac{ \langle \hat{w}_{\lambda} \rangle  - \langle \hat{w}_{-\lambda} \rangle }{2\lambda} + O(\lambda^2).
\end{equation}
We used $\lambda = 4 \cdot 10^{-6}$, and varied $\lambda$ to verify that $O(\lambda^2)$ effects are at least $\times 10$ smaller than the statistical error.

The plaquette results from the different models are shown in \Cref{fig:Wloopcompare}. As expected, increasing the model complexity leads to greater variance reduction. While this is just a selection of architectures and many improvements may be possible, it highlights the potential for further gains as model complexity and training effort are increased. 
We also evaluate the best model developed in this work on a larger $16^3 \times 32$ lattice geometry, which achieves a variance reduction of 53(1), compared to 54(4) at the $8^3 \times 32$ geometry. 
This confirms that the variance reduction is approximately volume-independent.

Glueball spectroscopy provides a more challenging application of the same learned flows.
Such calculations are remarkably difficult via standard methods, due to the signal-to-noise problem
~\cite{Athenodorou:2020ani,Abbott:2025irb,Urrea-Nino:2026cjj}, as highlighted in recent reviews~\cite{Vadacchino:2023vnc,Morningstar:2024vjk}.  
Here, we study the vacuum-subtracted glueball correlator,
$C(t) = \langle \mathcal G(t) \mathcal G(0) \rangle_{\rm c}$.

\pagebreak
Applying a time translation to the gauge fields, we insert $\mathcal{G}(t)$ on a different timeslice. We denote by $\lambda_t$ the source parameter that inserts the operator on a generic timeslice $t$. We then compute two-point functions as
\begin{equation}
     C(t) =  \frac{\langle \hat w_{\lambda_0 \lambda_t}  \rangle - \langle \hat w_{\lambda_0}  \rangle \langle \hat w_{ \lambda_t}  \rangle}{\lambda_0 \lambda_t} + O(\lambda). \label{eq:glue2pt}
\end{equation}
As before, $\lambda_t$ and $\lambda_0$ are made sufficiently small so that $O(\lambda)$ effects are negligible. The results are shown in \Cref{fig:results}; a reduction in the variance by more than three orders of magnitude is achieved, extending the signal by about four timeslices.

In a targeted glueball study, interpolating operators projected onto definite quantum numbers are typically used, in contrast to the choice of $\mathcal{G}(t)$ here. This can be done \emph{a posteriori} by combining the correlator studied here with those involving plaquette operators in other orientations, as discussed in the End Matter.

\ssection{Applications to QCD}
As a final demonstration, we compute the glueball correlation function in $N_f=2$ QCD. For this, we fine-tune the flow used in the Yang--Mills case with additional training at a small volume. The correlation function is also estimated using \Cref{eq:glue2pt}. The key difference is that evaluating the reweighting factors involves estimating the ratio of fermion determinants, for which we use the standard pseudofermion trick:
\begin{equation}
    \operatorname{det} \frac{D D^{\dagger}(V)}{D D^{\dagger}(U)}=\left\langle e^{-\phi^{\dagger}\left(M M^{\dagger}\right)^{-1} \phi} e^{\phi^{\dagger} \phi}\right\rangle_\phi,
\end{equation}
where $M = [D(U)]^{-1}D(V)$ in terms of the Dirac operator $D$, and $\langle \cdot \rangle_\phi$ denotes the average of Gaussian-distributed pseudofermions $\phi$. Since $D(U)$ and $D(V)$ differ only through the closely related gauge fields, a few pseudofermion samples lead to a low-variance estimate of the determinant ratio. To ensure variance reduction in \Cref{eq:glue2pt}, all reweighting factors must be evaluated with the same pseudofermion samples.

The results for the variance-reduced glueball correlation function in QCD are shown in \Cref{fig:results}. Over two orders of magnitude of variance reduction are achieved. 
This demonstrates the potential of the method to make studies of observables typically limited by gauge ensemble sizes tractable in the near future.

\ssection{Conclusion and outlook}
We have introduced a method that leverages the emerging technology of lattice normalizing flows within a systematically exact variance-improvement method. The results have demonstrated that the variance of expectation values of local gluonic operators can be reduced by orders of magnitude in both the Yang--Mills and QCD theories. Importantly, the learned generating-functional approach can provide estimators with variance that can be systematically driven towards zero by improving the model quality.

The numerical demonstrations correspond to medium-scale lattice QCD targets, and the results are stable under small variations of the lattice parameters. By training at smaller volumes than the target volume, we exploit transfer learning and benefit from the observed near volume-independence of the variance-reduction factor~\cite{Abbott:2026ylv}. This leaves no practical or theoretical obstacles to applying the method to state-of-the-art problems involving gluonic operators.
In contrast, we expect that less local insertions, such as larger Wilson loops or Wick-contracted quark operators, may be more challenging for our present architectures to learn.
This is strong motivation for future developments: we expect that there is still the opportunity for significant gains across many choices of observables by further technological advancement in the parametrization and training of flow models.

Potential applications of this work span many areas of strong-interaction physics, including hadron spectroscopy and structure, finite-temperature observables, and electromagnetic corrections to hadronic quantities.
As such, machine-learned flows are poised to become an indispensable tool in the lattice field theory repertoire.

\vspace{1cm}

\section*{Acknowledgements}

Special thanks to Denis Boyda and Julian Urban for extensive previous collaborations which led to ideas and code used in this work. We thank Michael Albergo, Pietro Butti, Kyle Cranmer, Mathis Gerdes, Sébastien Racanière, Alberto Ramos and Danilo J. Rezende for useful discussions. We are grateful to the Albert Einstein Center for Fundamental Physics and the Higgs Centre for Theoretical Physics for hosting GK and FRL for extended visits during which part of this work was carried out.

RA is an Ernest Kempton Adams Postdoctoral Fellow supported in part by the Ernest Kempton Adams fund for Physical Research of Columbia University and in part by U.S. DOE grant No. DE-SC0011941.
YF and PES are supported in part by the U.S.\ Department of Energy, Office of Science, Office of Nuclear Physics, under grant Contract Number DE-SC0011090. PES is additionally supported by the U.S.\ DOE Early Career Award DE-SC0021006, by a NEC research award, and by the Carl G and Shirley Sontheimer Research Fund. The work of FRL was supported in part by the Platform for Advanced Scientific Computing (PASC) project ``ALPENGLUE''. 
This document was prepared using the resources of the Fermi National Accelerator Laboratory (Fermilab), a U.S. Department of Energy, Office of Science, Office of High Energy Physics HEP User Facility. Fermilab is managed by Fermi Forward Discovery Group, LLC, acting under Contract No. 89243024CSC000002.
This work is supported by the U.S.\ National Science Foundation under Cooperative Agreement PHY-2019786 (The NSF AI Institute for Artificial Intelligence and Fundamental Interactions, \url{http://iaifi.org/}).
Part of the calculations were performed on \href{https://www.id.unibe.ch/hpc}{UBELIX}, the HPC cluster at the University of Bern.
Numerical experiments and data analysis used PyTorch~\cite{paszke2019pytorch}, NumPy~\cite{harris2020array}, and SciPy~\cite{2020SciPy-NMeth}. Figures were produced using matplotlib~\cite{Hunter:2007}.

\bibliographystyle{utphys}
\bibliography{main}

@article{Athenodorou:2020ani,
    author = "Athenodorou, Andreas and Teper, Michael",
    title = "{The glueball spectrum of SU(3) gauge theory in 3 + 1 dimensions}",
    eprint = "2007.06422",
    archivePrefix = "arXiv",
    primaryClass = "hep-lat",
    doi = "10.1007/JHEP11(2020)172",
    journal = "JHEP",
    volume = "11",
    pages = "172",
    year = "2020"
}

@inproceedings{Vadacchino:2023vnc,
    author = "Vadacchino, Davide",
    title = "{A review on Glueball hunting}",
    booktitle = "{39th International Symposium on Lattice Field Theory}",
    eprint = "2305.04869",
    archivePrefix = "arXiv",
    primaryClass = "hep-lat",
    month = "5",
    year = "2023"
}

@article{Morningstar:2024vjk,
    author = "Morningstar, Colin",
    title = "{Update on Glueballs}",
    eprint = "2502.02547",
    archivePrefix = "arXiv",
    primaryClass = "hep-lat",
    doi = "10.22323/1.466.0004",
    journal = "PoS",
    volume = "LATTICE2024",
    pages = "004",
    year = "2024"
}

@article{Urrea-Nino:2026cjj,
    author = "Urrea-Ni{\~n}o, Juan Andr{\'e}s and Knechtli, Francesco and Korzec, Tomasz and Peardon, Michael",
    title = "{Charmonium-Glueball spectroscopy with improved hadron creation operators}",
    eprint = "2603.20178",
    archivePrefix = "arXiv",
    primaryClass = "hep-lat",
    reportNumber = "WUB/26-00",
    month = "3",
    year = "2026"
}

@article{Necco:2001xg,
    author = "Necco, Silvia and Sommer, Rainer",
    title = "{The N(f) = 0 heavy quark potential from short to intermediate distances}",
    eprint = "hep-lat/0108008",
    archivePrefix = "arXiv",
    reportNumber = "DESY-01-095",
    doi = "10.1016/S0550-3213(01)00582-X",
    journal = "Nucl. Phys. B",
    volume = "622",
    pages = "328--346",
    year = "2002"
}

@article{Kitazawa:2016dsl,
    author = "Kitazawa, Masakiyo and Iritani, Takumi and Asakawa, Masayuki and Hatsuda, Tetsuo and Suzuki, Hiroshi",
    title = "{Equation of State for SU(3) Gauge Theory via the Energy-Momentum Tensor under Gradient Flow}",
    eprint = "1610.07810",
    archivePrefix = "arXiv",
    primaryClass = "hep-lat",
    reportNumber = "RIKEN-QHP-236, KYUSHU-HET-170, J-PARC-TH-0071",
    doi = "10.1103/PhysRevD.94.114512",
    journal = "Phys. Rev. D",
    volume = "94",
    number = "11",
    pages = "114512",
    year = "2016"
}

@article{Gockeler:2005rv,
    author = "Gockeler, M. and Horsley, R. and Irving, A. C. and Pleiter, D. and Rakow, P. E. L. and Schierholz, G. and Stuben, H.",
    title = "{A Determination of the Lambda parameter from full lattice QCD}",
    eprint = "hep-ph/0502212",
    archivePrefix = "arXiv",
    reportNumber = "DESY-05-028, EDINBURGH-2005-02, LTH-647, LU-ITP-2005-012",
    doi = "10.1103/PhysRevD.73.014513",
    journal = "Phys. Rev. D",
    volume = "73",
    pages = "014513",
    year = "2006"
}

@article{Tomiya:2022meu,
    author = "Tomiya, Akio and Terasaki, Satoshi",
    title = "{GomalizingFlow.jl: A Julia package for Flow-based sampling algorithm for lattice field theory}",
    eprint = "2208.08903",
    archivePrefix = "arXiv",
    primaryClass = "hep-lat",
    month = "8",
    year = "2022"
}

@article{Hermansson-Truedsson:2026zlp,
    author = "Hermansson-Truedsson, Nils and Kanwar, Gurtej",
    title = "{Normalizing flows for all-orders QED corrections in lattice field theory}",
    eprint = "2605.22444",
    archivePrefix = "arXiv",
    primaryClass = "hep-lat",
    month = "5",
    year = "2026"
}

@article{Abbott:2026ylv,
    author = "Abbott, Ryan and Boyda, Denis and Fu, Yang and Hackett, Daniel C. and Kanwar, Gurtej and Romero-L{\'o}pez, Fernando and Shanahan, Phiala E. and Urban, Julian M.",
    title = "{Variance reduction in lattice QCD observables via normalizing flows}",
    eprint = "2603.02984",
    archivePrefix = "arXiv",
    primaryClass = "hep-lat",
    reportNumber = "FERMILAB-PUB-26-0130-T, MIT-CTP/6010",
    month = "3",
    year = "2026"
}

@article{Abbott:2024mix,
    author = "Abbott, Ryan and Boyda, Denis and Hackett, Daniel C. and Kanwar, Gurtej and Romero-L{\'o}pez, Fernando and Shanahan, Phiala E. and Urban, Julian M. and Albergo, Michael S.",
    title = "{Practical applications of machine-learned flows on gauge fields}",
    eprint = "2404.11674",
    archivePrefix = "arXiv",
    primaryClass = "hep-lat",
    reportNumber = "FERMILAB-CONF-24-0007-T, MIT-CTP/5669",
    doi = "10.22323/1.453.0011",
    journal = "PoS",
    volume = "LATTICE2023",
    pages = "011",
    year = "2024"
}

@article{Bonanno:2025pdp,
    author = "Bonanno, Claudio and Bulgarelli, Andrea and Cellini, Elia and Nada, Alessandro and Panfalone, Dario and Vadacchino, Davide and Verzichelli, Lorenzo",
    title = "{Scaling flow-based approaches for topology sampling in $\mathrm{SU}(3)$ gauge theory}",
    eprint = "2510.25704",
    archivePrefix = "arXiv",
    primaryClass = "hep-lat",
    month = "10",
    year = "2025"
}

@article{Gerdes:2024rjk,
    author = "Gerdes, Mathis and de Haan, Pim and Bondesan, Roberto and Cheng, Miranda C. N.",
    title = "{Nonperturbative trivializing flows for lattice gauge theories}",
    eprint = "2410.13161",
    archivePrefix = "arXiv",
    primaryClass = "hep-lat",
    doi = "10.1103/31d5-hvp6",
    journal = "Phys. Rev. D",
    volume = "112",
    number = "9",
    pages = "094516",
    year = "2025"
}

@article{Abbott:2025irb,
    author = "Abbott, Ryan and Hackett, Daniel C. and Pefkou, Dimitra A. and Romero-L{\'o}pez, Fernando and Shanahan, Phiala E.",
    title = "{Lattice evidence that scalar glueballs are small}",
    eprint = "2508.21821",
    archivePrefix = "arXiv",
    primaryClass = "hep-lat",
    reportNumber = "MIT-CTP/5907, FERMILAB-PUB-25-0621-T, INT-PUB-25-021",
    month = "8",
    year = "2025"
}

@article{Catumba:2025ljd,
    author = "Catumba, Guilherme and Ramos, Alberto",
    title = "{Stochastic automatic differentiation and the signal to noise problem}",
    eprint = "2502.15570",
    archivePrefix = "arXiv",
    primaryClass = "hep-lat",
    doi = "10.1140/epjc/s10052-025-14690-0",
    journal = "Eur. Phys. J. C",
    volume = "85",
    number = "9",
    pages = "1037",
    year = "2025"
}

@article{Abbott:2024kfc,
    author = "Abbott, Ryan and Botev, Aleksandar and Boyda, Denis and Hackett, Daniel C. and Kanwar, Gurtej and Racani{\`e}re, S{\'e}bastien and Rezende, Danilo J. and Romero-L{\'o}pez, Fernando and Shanahan, Phiala E. and Urban, Julian M.",
    title = "{Applications of flow models to the generation of correlated lattice QCD ensembles}",
    eprint = "2401.10874",
    archivePrefix = "arXiv",
    primaryClass = "hep-lat",
    reportNumber = "MIT-CTP/5658, FERMILAB-PUB-24-0014-T",
    doi = "10.1103/PhysRevD.109.094514",
    journal = "Phys. Rev. D",
    volume = "109",
    number = "9",
    pages = "094514",
    year = "2024"
}

@article{Bacchio:2022vje,
    author = "Bacchio, Simone and Kessel, Pan and Schaefer, Stefan and Vaitl, Lorenz",
    title = "{Learning trivializing gradient flows for lattice gauge theories}",
    eprint = "2212.08469",
    archivePrefix = "arXiv",
    primaryClass = "hep-lat",
    doi = "10.1103/PhysRevD.107.L051504",
    journal = "Phys. Rev. D",
    volume = "107",
    number = "5",
    pages = "L051504",
    year = "2023"
}

@article{Bacchio:2023all,
    author = "Bacchio, Simone",
    title = "{A novel approach for computing gradients of physical observables}",
    eprint = "2305.07932",
    archivePrefix = "arXiv",
    primaryClass = "hep-lat",
    month = "5",
    year = "2023"
}

@article{Catumba:2023ulz,
    author = "Catumba, Guilherme and Ramos, Alberto and Zaldivar, Bryan",
    title = "{Stochastic automatic differentiation for Monte Carlo processes}",
    eprint = "2307.15406",
    archivePrefix = "arXiv",
    primaryClass = "hep-lat",
    month = "7",
    year = "2023"
}

@article{Albandea:2023wgd,
    author = "Albandea, David and Del Debbio, Luigi and Hern\'andez, Pilar and Kenway, Richard and Marsh Rossney, J. and Ramos, Alberto",
    title = "{Learning trivializing flows}",
    eprint = "2302.08408",
    archivePrefix = "arXiv",
    primaryClass = "hep-lat",
    reportNumber = "IFIC/23-07",
    doi = "10.1140/epjc/s10052-023-11838-8",
    journal = "Eur. Phys. J. C",
    volume = "83",
    number = "7",
    pages = "676",
    year = "2023"
}

@article{Bialas:2023fyj,
    author = "Bialas, Piotr and Korcyl, Piotr and Stebel, Tomasz",
    title = "{Training normalizing flows with computationally intensive target probability distributions}",
    eprint = "2308.13294",
    archivePrefix = "arXiv",
    primaryClass = "cs.LG",
    month = "8",
    year = "2023"
}

@article{Abbott:2023thq,
    author = {Ryan Abbott and Michael S. Albergo and Aleksandar Botev and Denis Boyda and Kyle Cranmer and Daniel C. Hackett and Gurtej Kanwar and Alexander G. D. G. Matthews and S\'ebastien Racani\`ere and Ali Razavi and Danilo J. Rezende and Fernando Romero-L\'opez and Phiala E. Shanahan and Julian M. Urban},
    title = "{Normalizing flows for lattice gauge theory in arbitrary space-time dimension}",
    eprint = "2305.02402",
    archivePrefix = "arXiv",
    primaryClass = "hep-lat",
    month = "5",
    year = "2023"
}

@article{Abbott:2022zsh,
    author = {Ryan Abbott and Michael S. Albergo and Aleksandar Botev and Denis Boyda and Kyle Cranmer and Daniel C. Hackett and Alexander G. D. G. Matthews and S\'ebastien Racani\`ere and Ali Razavi and Danilo J. Rezende and Fernando Romero-L\'opez and Phiala E. Shanahan and Julian M. Urban},
    title = "{Aspects of scaling and scalability for flow-based sampling of lattice QCD}",
    eprint = "2211.07541",
    archivePrefix = "arXiv",
    primaryClass = "hep-lat",
    reportNumber = "MIT-CTP/5496",
    doi = "10.1140/epja/s10050-023-01154-w",
    journal = "Eur. Phys. J. A",
    volume = "59",
    number = "11",
    pages = "257",
    year = "2023"
}

@article{paszke2019pytorch,
       author = {Paszke, Adam and others},
        title = "{PyTorch: An Imperative Style, High-Performance Deep Learning Library}",
      Cjournal = {arXiv e-prints},
         year = 2019,
        month = dec,
          eid = {arXiv:1912.01703},
        pages = {arXiv:1912.01703},
          doi = {10.48550/arXiv.1912.01703},
archivePrefix = {arXiv},
       eprint = {1912.01703},
 primaryClass = {cs.LG},
       adsurl = {https://ui.adsabs.harvard.edu/abs/2019arXiv191201703P}
}

@article{harris2020array,
  title={Array programming with NumPy},
  author={Harris, Charles R and Millman, K Jarrod and Van Der Walt, St{\'e}fan J and Gommers, Ralf and Virtanen, Pauli and Cournapeau, David and Wieser, Eric and Taylor, Julian and Berg, Sebastian and Smith, Nathaniel J and others},
  journal={Nature},
  volume={585},
  number={7825},
  pages={357--362},
  year={2020},
  publisher={Nature Publishing Group}
}

@article{2020SciPy-NMeth,
  title={SciPy 1.0: fundamental algorithms for scientific computing in Python},
  author={Virtanen, Pauli and Gommers, Ralf and Oliphant, Travis E and Haberland, Matt and Reddy, Tyler and Cournapeau, David and Burovski, Evgeni and Peterson, Pearu and Weckesser, Warren and Bright, Jonathan and others},
  journal={Nature methods},
  volume={17},
  number={3},
  pages={261--272},
  year={2020},
  publisher={Nature Publishing Group}
}

@article{Finkenrath:2022ogg,
    author = "Finkenrath, Jacob",
    title = "{Tackling critical slowing down using global correction steps with equivariant flows: the case of the Schwinger model}",
    eprint = "2201.02216",
    archivePrefix = "arXiv",
    primaryClass = "hep-lat",
    month = "1",
    year = "2022"
}

@article{DelDebbio:2021qwf,
    author = "Del Debbio, Luigi and Rossney, Joe Marsh and Wilson, Michael",
    title = "{Efficient Modelling of Trivializing Maps for Lattice $\phi^4$ Theory Using Normalizing Flows: A First Look at Scalability}",
    eprint = "2105.12481",
    archivePrefix = "arXiv",
    primaryClass = "hep-lat",
    month = "5",
    year = "2021"
}

@article{Nicoli:2020evf,
    author = {Nicoli, Kim A. and Nakajima, Shinichi and Strodthoff, Nils and Samek, Wojciech and M\"uller, Klaus-Robert and Kessel, Pan},
    title = "{Asymptotically unbiased estimation of physical observables with neural samplers}",
    eprint = "1910.13496",
    archivePrefix = "arXiv",
    primaryClass = "cond-mat.stat-mech",
    doi = "10.1103/PhysRevE.101.023304",
    journal = "Phys. Rev. E",
    volume = "101",
    number = "2",
    pages = "023304",
    year = "2020"
}

@inproceedings{Foreman:2021ljl,
    author = "Foreman, Sam and Izubuchi, Taku and Jin, Luchang and Jin, Xiao-Yong and Osborn, James C. and Tomiya, Akio",
    title = "{HMC with Normalizing Flows}",
    booktitle = "{38th International Symposium on Lattice Field Theory}",
    eprint = "2112.01586",
    archivePrefix = "arXiv",
    primaryClass = "cs.LG",
    month = "Dec",
    year = "2021"
}

@article{tabak2010,
author = "Tabak, Esteban G. and Vanden-Eijnden, Eric",
fjournal = "Communications in Mathematical Sciences",
journal = "Commun. Math. Sci.",
month = "03",
number = "1",
pages = "217--233",
publisher = "International Press of Boston",
title = "Density estimation by dual ascent of the log-likelihood",
url_dummy = "https://projecteuclid.org:443/euclid.cms/1266935020",
volume = "8",
year = "2010",
doi="10.4310/CMS.2010.v8.n1.a11"
}

@article{tabak2013,
author = {Tabak, E. G. and Turner, Cristina V.},
title = {A Family of Nonparametric Density Estimation Algorithms},
journal = {Communications on Pure and Applied Mathematics},
volume = {66},
number = {2},
pages = {145-164},
doi = {https://doi.org/10.1002/cpa.21423},
url_dummy = {https://onlinelibrary.wiley.com/doi/abs/10.1002/cpa.21423},
year = {2013}
}

@article{rezende2016variational,
      title={Variational Inference with Normalizing Flows}, 
      author={Danilo Jimenez Rezende and Shakir Mohamed},
      year={2016},
      eprint={1505.05770},
      archivePrefix={arXiv},
      primaryClass={stat.ML}
}

@article{Hunter:2007,
  Author    = {Hunter, J. D.},
  Title     = {Matplotlib: A 2D graphics environment},
  Journal   = {Computing in Science \& Engineering},
  Volume    = {9},
  Number    = {3},
  Pages     = {90--95},
  publisher = {IEEE COMPUTER SOC},
  doi       = {10.1109/MCSE.2007.55},
  year      = 2007
}

@article{Aliberti:2025beg,
    author = "Aliberti, R. and others",
    title = "{The anomalous magnetic moment of the muon in the Standard Model: an update}",
    eprint = "2505.21476",
    archivePrefix = "arXiv",
    primaryClass = "hep-ph",
    reportNumber = "CERN-TH-2025-101, FERMILAB-PUB-25-0344-T, INT-PUB-25-015, IPARCOS-UCM-25-029, KEK Preprint 2025-22, LTH 1403, MITP-25-037, UWThPh 2025-15, UWThPh
  2025-15, ZU-TH 37/25, IPARCOS-UCM-25-029",
    doi = "10.1016/j.physrep.2025.08.002",
    journal = "Phys. Rept.",
    volume = "1143",
    pages = "1--158",
    year = "2025"
}

@article{DallaBrida:2026kuo,
    author = {Dalla Brida, Mattia and H{\"o}llwieser, Roman and Knechtli, Francesco and Korzec, Tomasz and Ramos, Alberto and Sint, Stefan and Sommer, Rainer},
    title = "{High-precision calculation of the quark{\textendash}gluon coupling from lattice QCD}",
    doi = "10.1038/s41586-026-10339-4",
    journal = "Nature",
    volume = "652",
    number = "8109",
    pages = "328--334",
    year = "2026"
}

@article{Detmold:2021ulb,
    author = "Detmold, William and Kanwar, Gurtej and Lamm, Henry and Wagman, Michael L. and Warrington, Neill C.",
    title = "{Path integral contour deformations for observables in $SU(N)$ gauge theory}",
    eprint = "2101.12668",
    archivePrefix = "arXiv",
    primaryClass = "hep-lat",
    reportNumber = "FERMILAB-PUB-21-014-T, INT-PUB-21-002, MIT-CTP/5270",
    doi = "10.1103/PhysRevD.103.094517",
    journal = "Phys. Rev. D",
    volume = "103",
    number = "9",
    pages = "094517",
    year = "2021"
}

@article{Detmold:2020ncp,
    author = "Detmold, William and Kanwar, Gurtej and Wagman, Michael L. and Warrington, Neill C.",
    title = "{Path integral contour deformations for noisy observables}",
    eprint = "2003.05914",
    archivePrefix = "arXiv",
    primaryClass = "hep-lat",
    reportNumber = "FERMILAB-PUB-20-095-T, INT-PUB-20-007, MIT-CTP/5182",
    doi = "10.1103/PhysRevD.102.014514",
    journal = "Phys. Rev. D",
    volume = "102",
    number = "1",
    pages = "014514",
    year = "2020"
}

@article{Lin:2023svo,
    author = "Lin, Yin and Detmold, William and Kanwar, Gurtej and Shanahan, Phiala E. and Wagman, Michael L.",
    title = "{Signal-to-noise improvement through neural network contour deformations for 3D {\ensuremath{\boldsymbol{\mathit{S}}}}{\ensuremath{\boldsymbol{\mathit{U}}}}(2) lattice gauge theory}",
    eprint = "2309.00600",
    archivePrefix = "arXiv",
    primaryClass = "hep-lat",
    reportNumber = "FERMILAB-CONF-23-485-T",
    doi = "10.22323/1.453.0043",
    journal = "PoS",
    volume = "LATTICE2023",
    pages = "043",
    year = "2024"
}

@article{Lawrence:2023sfc,
    author = "Lawrence, Scott and Yamauchi, Yukari",
    title = "{Convex optimization of contour deformations}",
    eprint = "2311.13002",
    archivePrefix = "arXiv",
    primaryClass = "hep-lat",
    doi = "10.1103/PhysRevD.110.014508",
    journal = "Phys. Rev. D",
    volume = "110",
    number = "1",
    pages = "014508",
    year = "2024"
}

@article{Bhattacharya:2023pxx,
    author = "Bhattacharya, Tanmoy and Lawrence, Scott and Yoo, Jun-Sik",
    title = "{Control variates for lattice field theory}",
    eprint = "2307.14950",
    archivePrefix = "arXiv",
    primaryClass = "hep-lat",
    doi = "10.1103/PhysRevD.109.L031505",
    journal = "Phys. Rev. D",
    volume = "109",
    number = "3",
    pages = "L031505",
    year = "2024"
}

@article{Oh:2025fpq,
    author = "Oh, Hyunwoo",
    title = "{Training neural control variates using correlated configurations}",
    eprint = "2505.07719",
    archivePrefix = "arXiv",
    primaryClass = "hep-lat",
    doi = "10.1103/kpc8-cj7p",
    journal = "Phys. Rev. D",
    volume = "112",
    number = "7",
    pages = "074501",
    year = "2025"
}

@article{Lawrence:2024xsi,
    author = "Lawrence, Scott",
    title = "{Schwinger-Dyson control variates for lattice fermions}",
    eprint = "2404.10707",
    archivePrefix = "arXiv",
    primaryClass = "hep-lat",
    month = "4",
    year = "2024"
}

@article{Bedaque:2023ovz,
    author = "Bedaque, Paulo F. and Oh, Hyunwoo",
    title = "{Leveraging neural control variates for enhanced precision in lattice field theory}",
    eprint = "2312.08228",
    archivePrefix = "arXiv",
    primaryClass = "hep-lat",
    doi = "10.1103/PhysRevD.109.094519",
    journal = "Phys. Rev. D",
    volume = "109",
    number = "9",
    pages = "094519",
    year = "2024"
}

@article{Yoon:2018krb,
    author = "Yoon, Boram and Bhattacharya, Tanmoy and Gupta, Rajan",
    title = "{Machine Learning Estimators for Lattice QCD Observables}",
    eprint = "1807.05971",
    archivePrefix = "arXiv",
    primaryClass = "hep-lat",
    reportNumber = "LA-UR-18-26411",
    doi = "10.1103/PhysRevD.100.014504",
    journal = "Phys. Rev. D",
    volume = "100",
    number = "1",
    pages = "014504",
    year = "2019"
}

@article{Parisi:1983ae,
    author = "Parisi, G.",
    editor = "Itzykson, C. and Pomeau, Y. and Sourlas, N.",
    title = "{The Strategy for Computing the Hadronic Mass Spectrum}",
    reportNumber = "LNF-83-36-P",
    doi = "10.1016/0370-1573(84)90081-4",
    journal = "Phys. Rept.",
    volume = "103",
    pages = "203--211",
    year = "1984"
}

@inproceedings{Lepage:1989hd,
    author = "Lepage, G. Peter",
    title = "{The Analysis of Algorithms for Lattice Field Theory}",
    booktitle = "{Theoretical Advanced Study Institute in Elementary Particle Physics}",
    reportNumber = "CLNS-89-971",
    month = "6",
    year = "1989"
}

@inproceedings{Albergo:2026fwx,
    author = "Albergo, Michael S. and Kanwar, Gurtej",
    title = "{A Monte Carlo estimator of flow fields for sampling and noise problems}",
    booktitle = "{42th International Symposium on Lattice Field Theory}",
    eprint = "2603.00252",
    archivePrefix = "arXiv",
    primaryClass = "hep-lat",
    month = "2",
    year = "2026"
}

@article{brenier1991polar,
  title={Polar factorization and monotone rearrangement of vector-valued functions},
  author={Brenier, Yann},
  journal={Communications on pure and applied mathematics},
  volume={44},
  number={4},
  pages={375--417},
  year={1991},
  publisher={Wiley Online Library}
}

\appendix
\crefalias{section}{appendix}

\clearpage

\begin{widetext}

\begin{center}
    \large\textbf{End Matter}
\end{center}
\vspace{-2em}

\section{Multilinear source expansion and exact control variates}
The main text involves the parameterized family of optimal flows $f_{\bs{\lambda}}$, but we are ultimately only interested in their derivatives at $\bs{\lambda} = 0$. Here we simplify this high-dimensional parameterization problem by expanding up to linear order in all source coefficients $\lambda_1, \dots, \lambda_n$. The result is an exact control variate for the $n$-point correlation function. 

In order to preserve the Lie group structure, we first write and expand the flow as
\begin{equation} \label{eq:flow-expansion}
    V_{\bs\lambda} \equiv f_{\bs{\lambda}}(U) = \exp(A_{\bs{\lambda}}(U)) \, U
    = \exp\Big[ \sum_i \lambda_i F_{\{ i\}}(U) + \sum_{i < j} \lambda_i \lambda_j F_{\{i,j\}}(U) + \dots \Big] \, U,
\end{equation}
where $A_{\bs{\lambda}}$ is algebra-valued, and terms which are quadratic in at least one $\lambda_i$ are omitted here and below. 
We introduce the notation $B \subseteq [n] \equiv \{1, \dots, n\}$ for subsets of the $n$ indices;
we define $\lambda_B \equiv \prod_{i \in B} \lambda_i$, $\partial_B \equiv \prod_{i \in B} \partial/\partial \lambda_i$, and ${\rm OP}(B)$ ($\Pi(B)$) for the set of ordered (unordered) partitions \(P=(P_1,\ldots,P_m)\) of nonempty, pairwise-disjoint subsets with union \(B\). Then the needed terms of the expansion can be compactly written as
\begin{equation}
A_{\bs\lambda} = \sum_{\emptyset \neq B \subseteq [n]} \lambda_B F_B(U).
\end{equation}
Only these multilinear terms can contribute to the derivative at $\bs{\lambda} = 0$, so the problem of parameterizing the family of $\bs{\lambda}$-dependent flows reduces to the problem of parameterizing the $2^n - 1$ algebra-valued, $\bs{\lambda}$-independent \emph{flow fields} $F_B(U)$.

To define an optimization objective on the flow fields $F_B$, we next expand the terms of the (log) weight $\log \hat{w}_{\bs \lambda} = \bs\lambda \cdot \bs{\mathcal{O}} - [S(V_{\bs\lambda}) - S(U)] + \ln |\det J_f|$.
The first term is already in appropriate series form. The second involves a Lie-Taylor series of the action,
\begin{equation} \label{eq:action-expansion}
    S(V_{\bs\lambda}) - S(U) = \sum_{k=1}^\infty \frac{1}{k!} \Big[ \sum_{\emptyset \neq B \subseteq [n]} \lambda_B F_B(U) \cdot \nabla \Big]^k S(U)
    = \sum_{\emptyset \neq B \subseteq [n]} \lambda_B \sum_{P \in {\rm OP}(B)} \frac{1}{|P|!} (F_{P_1} \cdot \nabla) \dots (F_{P_{|P|}} \cdot \nabla) \, S(U),
\end{equation}
where $\nabla$ indicates the algebra-valued Lie derivative with respect to the $U$ dependence in $S(U)$ and $(X \cdot \nabla) \equiv X^a \nabla^a$ in terms of the algebra-projected coefficients $X^a = -2 \Tr[T^a X]$; the anti-Hermitian generators are normalized as $\Tr[T^a T^b] = -\tfrac{1}{2} \delta^{ab}$. Finally, we expand the third term, starting with expanding the exponential in \Cref{eq:flow-expansion},
\begin{equation} \label{eq:flow-expansion-2}
    V_{\bs \lambda} = \Big[
    \sum_{k=0}^{\infty} \frac{1}{k!} (\sum_{\emptyset \neq B \subseteq [n]} \lambda_B F_B(U))^k
    \Big] \, U
    =
    \sum_{B \subseteq [n]} \lambda_B \Big[\sum_{P \in {\rm OP}(B)} \frac{1}{|P|!} F_{P_1} \dots F_{P_{|P|}} \Big] \, U.
\end{equation}
We denote the term in square brackets $H_B(U) \equiv [\dots]$.
Then the Jacobian can be expanded as,
\begin{equation} \label{eq:jac-expansion}
\begin{aligned}
    J^{ab}_f &=  \delta^{ab} + \sum_{\emptyset \neq B \subseteq [n]} \lambda_B \Bigg\{ - 2 \mathrm{Re}\mathrm{Tr}[T^a (\nabla^b H_B + 2 H_B T^b)] 
    - \sum_{\substack{P \in \mathrm{OP}(B) \\ |P|=2}} 2\mathrm{Re}\mathrm{Tr}[T^a (\nabla^b H_{P_1} + H_{P_1} T^b) H_{P_2}] \Bigg\}.
\end{aligned}
\end{equation}
Defining the quantity in curly braces as a matrix in the space of algebra coefficients, $X_B^{ab} \equiv \{ \dots \}$, and using $\ln |\det J_f| = \tr \ln J_f$, this term expands as
\begin{equation} \label{eq:log-jac-expansion}
\begin{aligned}
    \ln |\det J_f| 
    &= \tr \left(
    \sum_{k=1}^{\infty} \frac{(-1)^{k-1}}{k} \big[\sum_{\emptyset \neq B \subseteq [n]} \lambda_B X_B \big]^k
    \right)
    = \sum_{\emptyset \neq B \subseteq [n]} \lambda_B \sum_{P \in \mathrm{OP}(B)} \frac{(-1)^{|P|-1}}{|P|} \tr \left( X_{P_1} \dots X_{P_{|P|}} \right).
\end{aligned}
\end{equation}
Combining the terms and definitions in \Cref{eq:flow-expansion-2,eq:action-expansion,eq:jac-expansion,eq:log-jac-expansion}, the result is, up to linear order in all parameters,
\begin{align}
    \log \hat{w}_{\bs\lambda} &= \sum_{\emptyset \neq B \subseteq [n]} \lambda_B C_B(U), \label{eq:logw-expansion} \\
    C_B(U) &\equiv \sum_{P \in \mathrm{OP}(B)} \left[ \frac{(-1)^{|P|-1}}{|P|} \tr(X_{P_1} \dots X_{P_{|P|}}) - \frac{1}{|P|!} (F_{P_1} \cdot \nabla) \dots (F_{P_{|P|}} \cdot \nabla) \, S(U)\right]  \; + \begin{cases}
        \mathcal{O}_i, & B = \{i\} \\
        0, & \text{otherwise}.
    \end{cases}
\end{align}
All $\bs\lambda$ dependence is carried by the series and all terms $C_B(U)$ depend only on the untransformed input $U$. 
For the following step, we also define $\mathcal{O}_{B} \equiv \prod_{i \in B} \mathcal{O}_i$.

For the optimal flow, we must have $\log \hat{w}_{\bs\lambda} = \log (Z_{\bs\lambda}/Z_0)$. Expanding $Z_{\bs\lambda}/Z_0 = \left< \exp(\bs\lambda \cdot \bs{\mathcal{O}}) \right>$ and equating the two series gives the optimality conditions for the flow fields,
\begin{equation} \label{eq:optimal-expanded}
    C_B(U) \xrightarrow{\substack{\mathrm{perfect} \\ \mathrm{flow}}}  \sum_{P \in \mathrm{OP}(B)} \frac{(-1)^{|P|-1}}{|P|} \left< \mathcal{O}_{P_1} \right> \dots \left< \mathcal{O}_{P_{|P|}} \right>,
    \quad \forall B \subseteq [n].
\end{equation}
The existence of a solution for $f_{\bs\lambda}$ implies the solvability of \Cref{eq:optimal-expanded}. As argued in the one-operator case in Ref.~\cite{Abbott:2026ylv}, the mean-squared-error of these equations can be used as a Physics-Informed Neural Network (PINN) loss function to directly optimize the flow fields $F_B$.

The series expansion in \Cref{eq:logw-expansion} also allows us to define the exact estimator directly at $\bs\lambda = 0$ by evaluating the source derivatives in Eq.~(5) of the main text. We have:
\begin{equation}
    \frac{\partial^n}{\partial \lambda_1 \dots \partial \lambda_n} \hat{w}_{\bs\lambda}
    = \sum_{P \in \Pi([n])} (\partial_{P_1} \log \hat{w}_{\bs\lambda}) \dots (\partial_{P_{|P|}} \log \hat{w}_{\bs\lambda}) 
    = \boxed{\sum_{P \in \Pi([n])} C_{P_1}(U) \dots C_{P_{|P|}}(U)}.
\end{equation}
By the analysis in the main text, the expectation value is exact, $\left< \mathcal{O}_1 \dots \mathcal{O}_n \right> = \sum_P \left< C_{P_1} \dots C_{P_{|P|}} \right>$, and the difference gives a zero-mean control variate. It can be verified that inserting \Cref{eq:optimal-expanded} into this estimator gives the zero-variance estimator equal to $\left< \mathcal{O}_1 \dots \mathcal{O}_n \right>$ for all inputs $U$.

\section{Factorized flows and symmetry-related operators}
\label{app:factorized-flows}

In the main text, a flow was trained to insert a timeslice-averaged plaquette operator, ${G_{\mu\nu}(t)=\sum_{\vec x}\operatorname{Re}\operatorname{Tr}
P_{\mu\nu}(\vec x,t)}$, for the definite orientation \(G_{12}\). Using the factorized approach, the same trained flow can also be reused for operators related by lattice symmetries. Suppose \(R\) is a spatial lattice rotation or reflection that preserves the action, \(S(RU)=S(U)\), and maps one plaquette orientation into another, \(G_{\rho\sigma}[U]=G_{\mu\nu}[RU]\).
If \(f^{\mu\nu}_{\lambda}\) is trained to insert \(G_{\mu\nu}\), then
\begin{equation}
    f^{\rho\sigma}_{\lambda}
  =R^{-1}\circ f^{\mu\nu}_{\lambda}\circ R
\end{equation}
inserts \(G_{\rho\sigma}\) and yields the same variance reduction for the corresponding one-point function. For example, if \(R_{23}\) exchanges the 2- and 3-directions so that \(G_{13}[U]=G_{12}[R_{23}U]\), then \(f^{13}_{\lambda}=R_{23}^{-1}\circ f^{12}_{\lambda}\circ R_{23}\) inserts \(G_{13}\).

For two-point functions, we likewise expect approximately the same variance reduction. To illustrate this, we apply a factorized flow to estimate the connected correlation functions
\begin{equation}
  C_{12,12}(t)
  = \left\langle G_{12}(t)G_{12}(0)\right\rangle_c,
  \qquad
  C_{12,13}(t)
  = \left\langle G_{12}(t)G_{13}(0)\right\rangle_c.
\end{equation}
The results are shown in \Cref{fig:rotation}, where the two correlators exhibit similar variance reduction factors.
Correlators with definite cubic-group quantum numbers can then be assembled as linear combinations of correlators built from operators with fixed orientations.

\begin{figure}[b]
    \centering
    \includegraphics{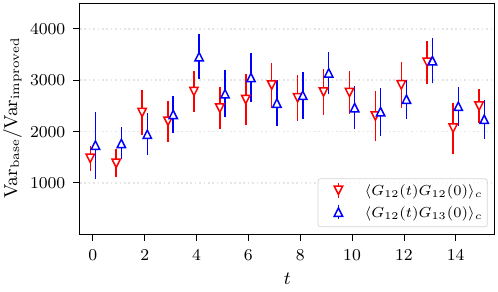}
    \caption{ Achieved variance reduction for two glueball correlation functions with operators in different orientations, evaluated using the same trained flow.  
    Results are shown for Yang-Mills at $\beta=6.0$, using an $8^3 \times 32$ lattice geometry. 
    }
    \label{fig:rotation}
\end{figure}

\end{widetext}
\end{document}